\documentclass[a4paper,11pt]{article}
\usepackage{pos}

\title{Improved topological sampling for lattice gauge theories}

\author*[a]{David Albandea}
\author[a]{Pilar Hernández}
\author[a]{Alberto Ramos}
\author[a,b]{Fernando Romero-López}

\affiliation[a]{Instituto de Física Corpuscular (CSIC -- University of Valencia),\\
  Parque Científico, C/Catedrático José Beltrán, 2, 46980, Paterna, Valencia,
  Spain}
\affiliation[b]{Center for Theoretical Physics, Massachusetts Institute of Technology,\\
Cambridge, MA 02139, USA}

\emailAdd{david.albandea@uv.es}
\emailAdd{m.pilar.hernandez@uv.es}
\emailAdd{alberto.ramos@ific.uv.es}
\emailAdd{fernando.romero@uv.es}

\abstract{
	Standard sampling algorithms for lattice QCD suffer from topology
	freezing (or critical slowing down) when approaching the continuum
	limit, thus leading to poor sampling of the distinct topological
	sectors. I will present a modified Hamiltonian Monte Carlo (HMC)
	algorithm that triggers topological sector jumps during the assembly of
	Markov chain of lattice configurations. We study its performance in the
	2D Schwinger model and compare it to alternative methods, such as fixing
	topology or master field. We then briefly discuss the difficulties of the
	algorithm in a SU(2) gauge model in 4D.
}

\FullConference{%
 The 38th International Symposium on Lattice Field Theory, LATTICE2021
  26th-30th July, 2021
  Zoom/Gather@Massachusetts Institute of Technology \\
  Report number: IFIC/21-41 \\
  Report number: MIT-CTP/5346
}

\begin{document}
\maketitle

\section{Introduction}

\subsection{Topology freezing}

\begin{figure*}[t]
	\centering
\includegraphics[width=12cm,keepaspectratio]{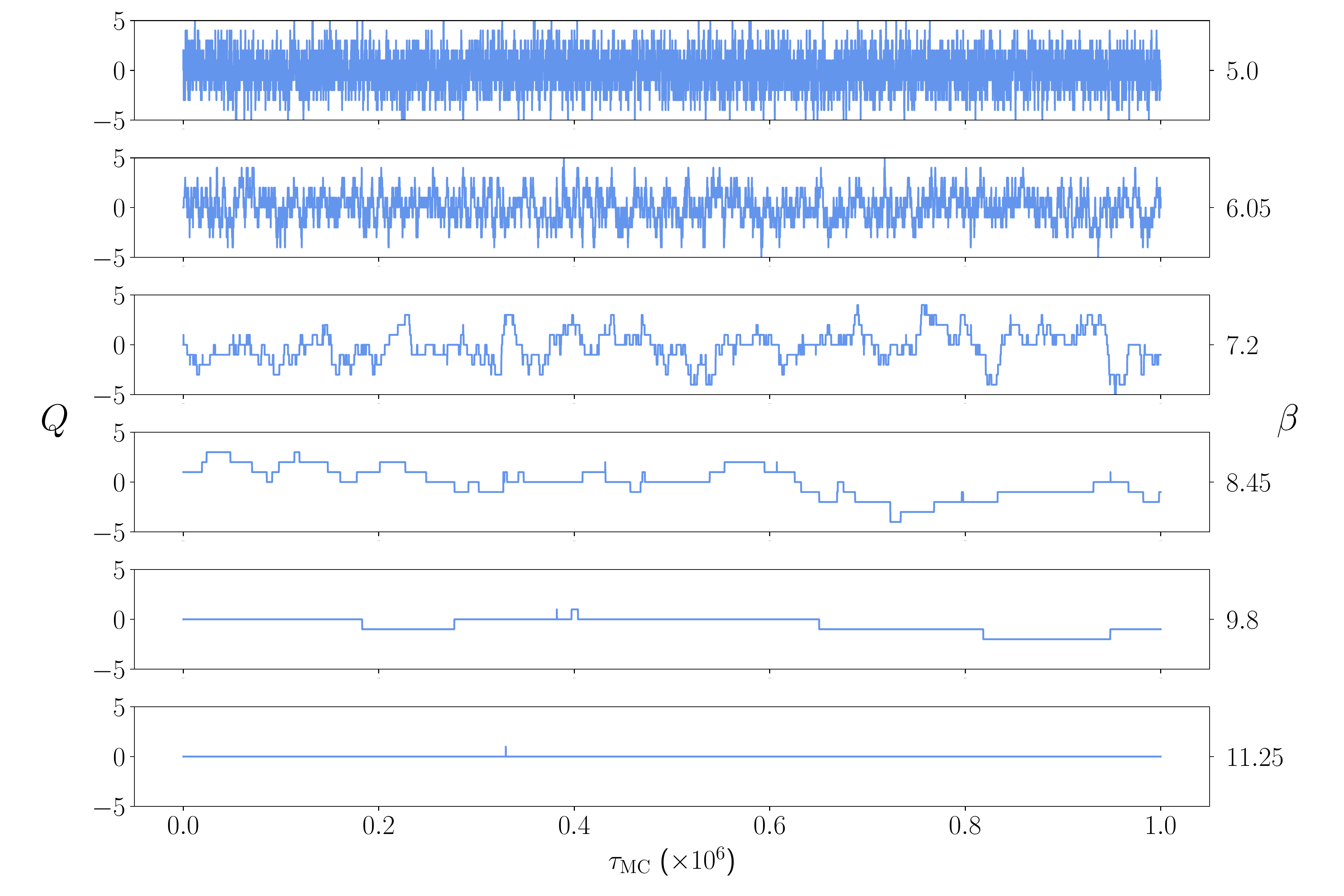}	
	\caption{Monte Carlo history of the topological charge $Q$ for increasing values of $\beta$ in a Markov chain of $10^6$ HMC configurations.}
	\label{fig:hmc_freeze}
\end{figure*}

		

Fig. \ref{fig:hmc_freeze} shows the typical scenario in lattice simulations
concerning topological observables: when we look at the Markov Chain Monte Carlo
history of the topological charge $Q$ at rough lattice spacings as in the upper part
of the plot, it fluctuates nicely;  however, when we go to finer lattice
spacings to take the continuum limit the topological charge freezes and one can
only take samples from one or two topological sectors, leading to long
autocorrelation times.

This happens because in the continuum limit the different topological sectors
are separated by barriers of infinite action, or, equivalently, by wide regions
in sampling space where the probability distribution $p(S)=e^{-S[ U ]}$ is zero,
causing update algorithms such as HMC to find it increasingly difficult to cross
from a topological sector to another and therefore always propose configurations
with the same topological charge $Q$.

The motivation of this work was to build a new algorithm that proposes
transitions to different topological sectors more frequently than HMC does by
doing a transformation over the gauge links \cite{Albandea_2021}.

\subsection{Schwinger model}

Of course, the final goal is to build the algorithm for QCD, but we started out
with a U(1) gauge theory in 2D and studied the pure gauge case and also the case
with $N_{f} = 2$ dynamical fermions. The Wilson lattice formulation of the U(1)
gauge theory is
\begin{eqnarray}
Z = \int \prod_{l} d U_l ~e^{-S_p[U]} \equiv \int \prod_{l} d U_l  ~e^{{\beta\over 2} \sum_p U_p+U_p^\dagger},
\end{eqnarray}
where $U_l$ and $U_p$ are the standard link and $1 \times 1$ Wilson loop,
respectively. This model is usually treated as a benchmark in the community,
also en machine learning \cite{Kanwar_U1} and tensor network approaches (see
\cite{Banuls_TN} for a review), because it is very similar to QCD: beyond being
a much simpler theory, both of them have topology and a mass gap (with $N_{f}=2$);
also, for $N_{f}=0$ there are analytical results, even at finite volume and for
all $\beta$ \cite{Kovacs_1995,Bonati_2019_1,Bonati_2019_2}, which is important
to check our results; and finally, the topological charge in this model has a
geometrical definition and it is exactly an integer,
\begin{align}
	Q \equiv \frac{-i}{2\pi}\sum_{p}^{} \ln U_{p} \in \mathbb{Z}.
\end{align}

\subsection{Metropolis--Hastings algorithm}

The idea to construct the new algorithm is to modify the Hybrid Monte Carlo
algorithm, which is a particular case of a Metropolis--Hastings algorithm
\cite{MH_1970}, whose
key ingredients are: 
\begin{enumerate}
	\item
		The target distribution from which we want to get samples; in
		our case, $p(U) = e^{-S[ U ]}$.
	\item
		A proposal distribution $q(U' | U)$, which is used to propose a new
		state in the Markov Chain from the configuration $U$. In HMC
		these are the Hamilton equations of motion.
	\item
		The accept-reject step, where we decide if we accept the
		proposed configuration or not, with probability
		\begin{align}
			p_{\text{acc}}(U'|U) = \min \left\{ 1,
			\frac{p(U')}{p(U)} \right\}.
			\label{eq:MHacc}
		\end{align}
\end{enumerate}
The performance of the algorithm will essentially be given by how efficiently
the proposal distribution $q(U'|U)$ is able to propose configurations from the relevant
regions in the sampling space. With this in mind, we built up a transformation
that made transitions between topological sectors.

\section{Winding Hybrid Monte Carlo}

\subsection{Winding transformation}

\begin{figure*}[t]
	\centering
\includegraphics[width=7cm,keepaspectratio]{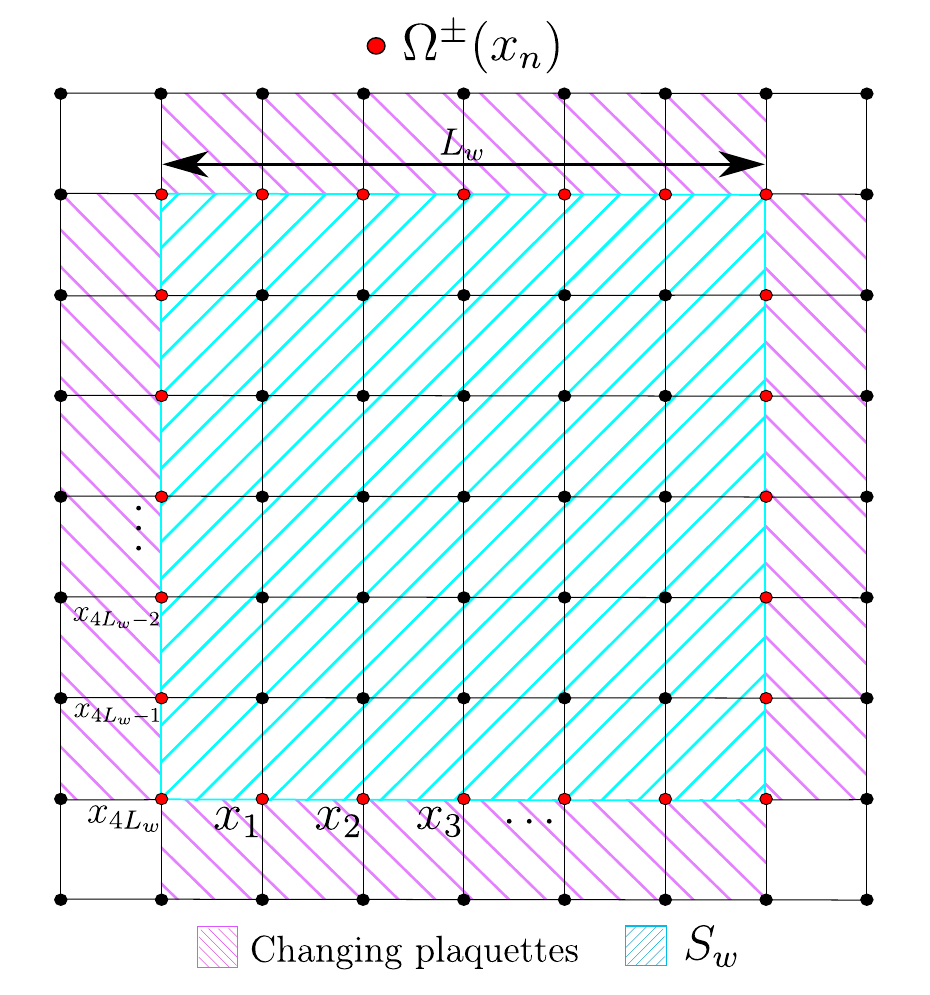}	
\includegraphics[width=8cm,keepaspectratio]{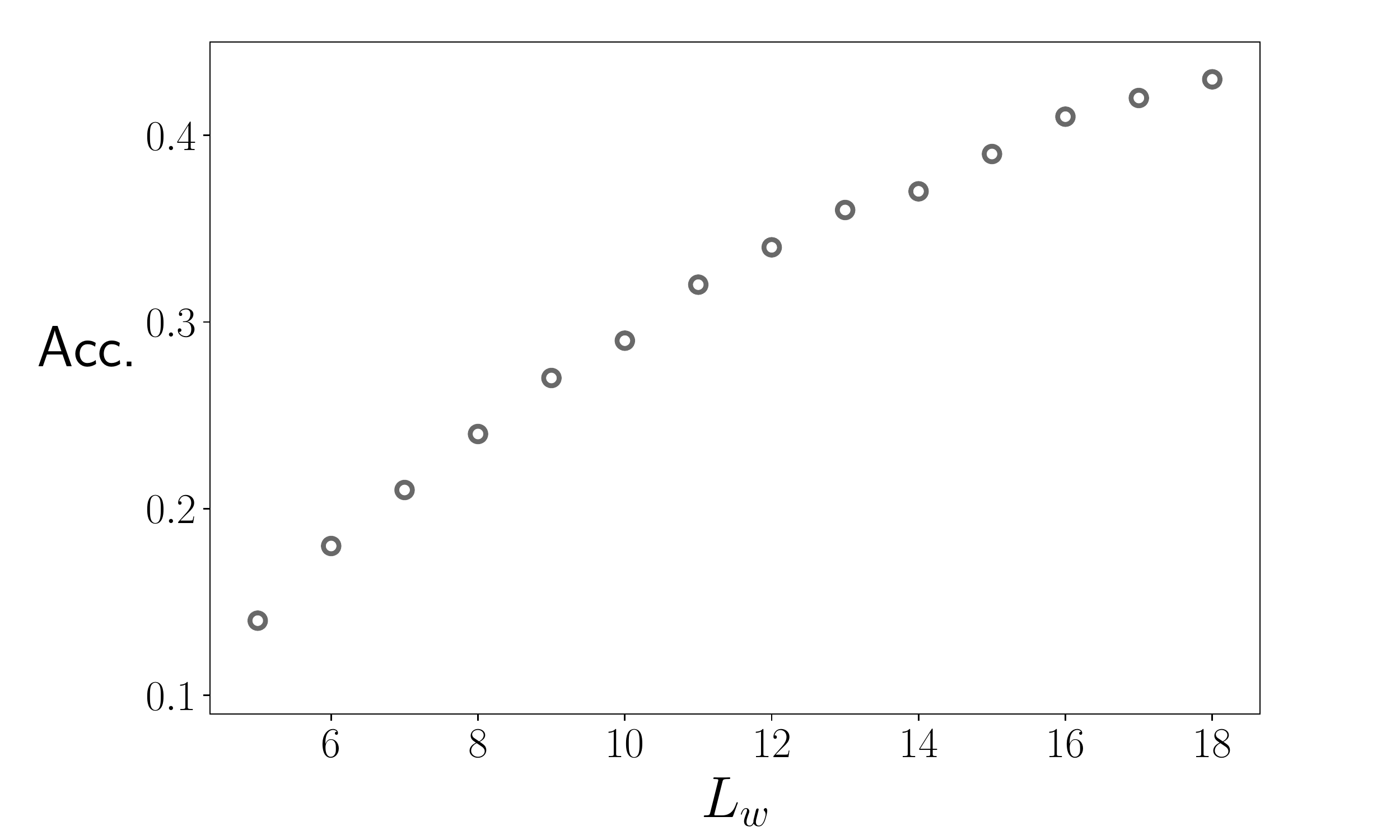}	
\caption{(Left) Sketch of a winding transformation of size $L_{w}$. Gauge links
in the blue region transform according to Eq. \eqref{eq:wtrafo}, while the
others stay the same. (Right) Dependence on the winding size $L_{w}$ of the
acceptance of a winding step at $\beta=5$.}
	\label{fig:winding_trafo}
\end{figure*}

The transformation that will trigger jumps between topological sectors is
defined as
\begin{align}
	U_{\mu}(x) \rightarrow U_{\mu}^{\Omega}(x) \equiv
	\Omega(x)U_{\mu}(x)\Omega^\dagger(x+\hat{\mu}) \qquad \text{if both }
	x,\ x+\hat{\mu} \in S_{w}.
	\label{eq:wtrafo}
\end{align}
This looks like a gauge transformation, but the key point is that we only
perform the transformation if the links at $x$ and $x + \hat{\mu}$ are both in a
region $S_{w}$ of size $L_{w}\times L_{w}$, depicted in blue in Fig.
\ref{fig:winding_trafo} (left). There, the plaquettes in the blue region will not
change, but the ones in violet will be affected because we are not completing
the gauge transformation outside the blue region. Finally the field $\Omega(x_{n})$ is
defined on the boundary of $S_{w}$ (red points) and is constructed such that the
violet plaquettes will change the topological charge of the configuration in one
unit,
\begin{align}
	\Omega^{\pm}(x_{n}) = e^{\pm i \frac{\pi}{2} \frac{n}{L_{w}}},
\end{align}
where the $+$ sign defines a winding and the $-$ an antiwinding. The sign is
chosen with 50\% probability and is common for the $n$ points, ensuring that the
transformation will yield a change in the topological charge of $\Delta Q = \pm
1$. Therefore, there is only one free parameter in this transformation, $L_{w}$,
and it will need to be tuned to optimize the acceptance of the algorithm. 

\subsection{Winding Hybrid Monte Carlo (wHMC)}

With this, we define the winding-step transition probability, 
\begin{align}
T(U \rightarrow U') = \frac{1}{2} \delta(U' - U^{\Omega^{+}}) + \frac{1}{2}
\delta(U' - U^{\Omega^{-}}) 
\end{align}
where we perform a winding or antiwinding transformation with 50\% probability,
which along with the accept-reject step of Eq. \eqref{eq:MHacc} can be shown to
satisfy detailed balance \cite{Albandea_2021}; and we combine this transformation
with HMC to have ergodicity, building a new algorithm which we call winding HMC
(wHMC), whose structure is:
\begin{enumerate}
	\item
		Perform a molecular dynamics evolution using HMC.
	\item
		Accept or reject the new configuration using Eq.
		\eqref{eq:MHacc}.
	\item
		Perform a winding or antiwinding transformation. 
	\item
		Accept or reject the new configuration using Eq.
		\eqref{eq:MHacc}.
	\item
		Repeat.
\end{enumerate}
This defines a wHMC step. The first thing to check is whether this new algorithm
has acceptance in the pure gage theory and how it depends on the size of the
transformation, and one can show that the mean variation of the action goes to
zero as one increases the size of the winding region $L_{w}$,
\begin{align}
	\left< \Delta S \right> \approx \frac{\beta \pi^2}{2L_{w}}.
\end{align}
Therefore we expect that the acceptance will grow with the size of the winding
up to a maximum of 50\%, as one can see in Fig. \ref{fig:winding_trafo} (right).

\section{$N_{f} = 0$ results}

\begin{figure*}[t]
	\centering
\includegraphics[width=7.5cm,keepaspectratio]{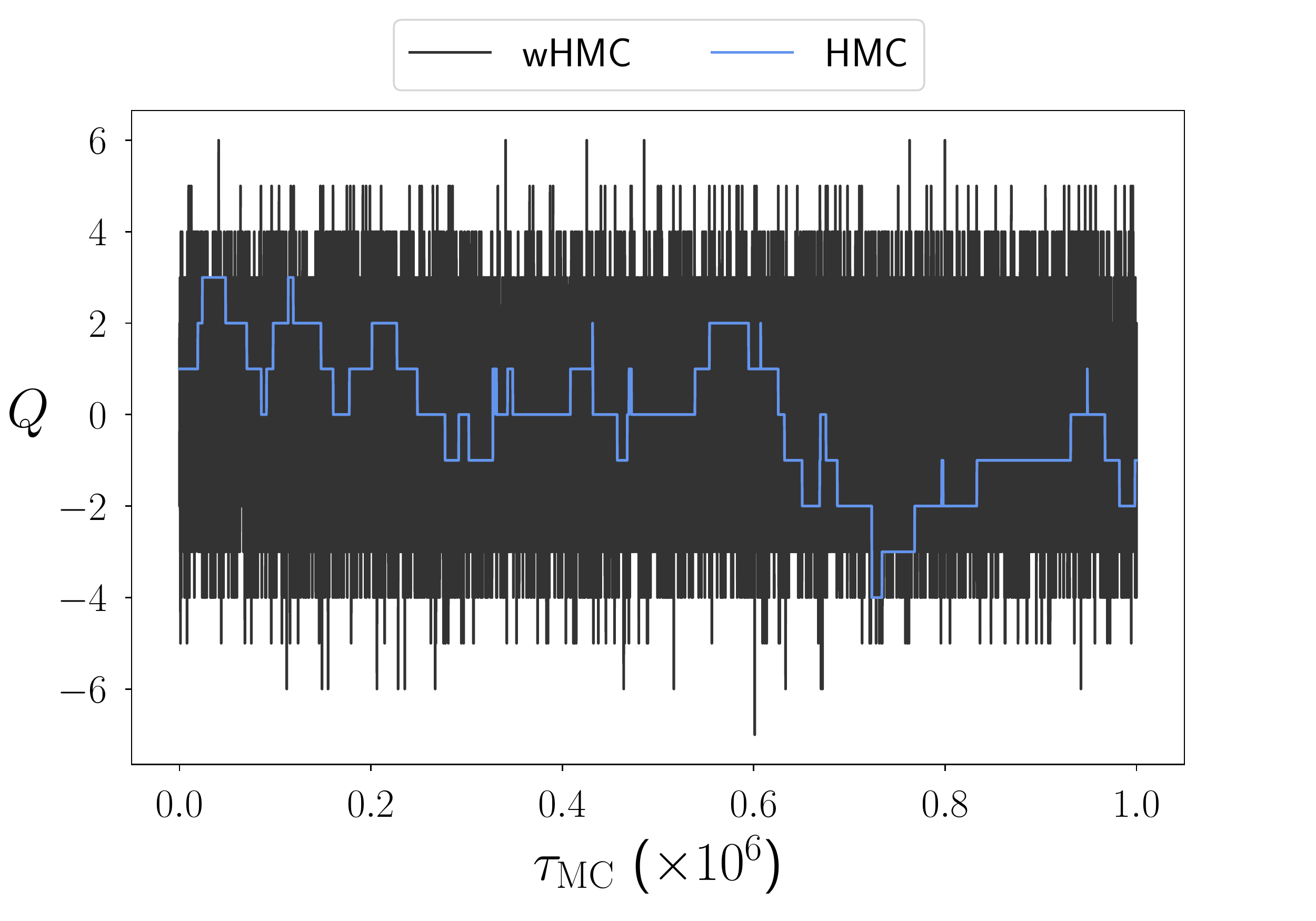}	
\includegraphics[width=7.5cm,keepaspectratio]{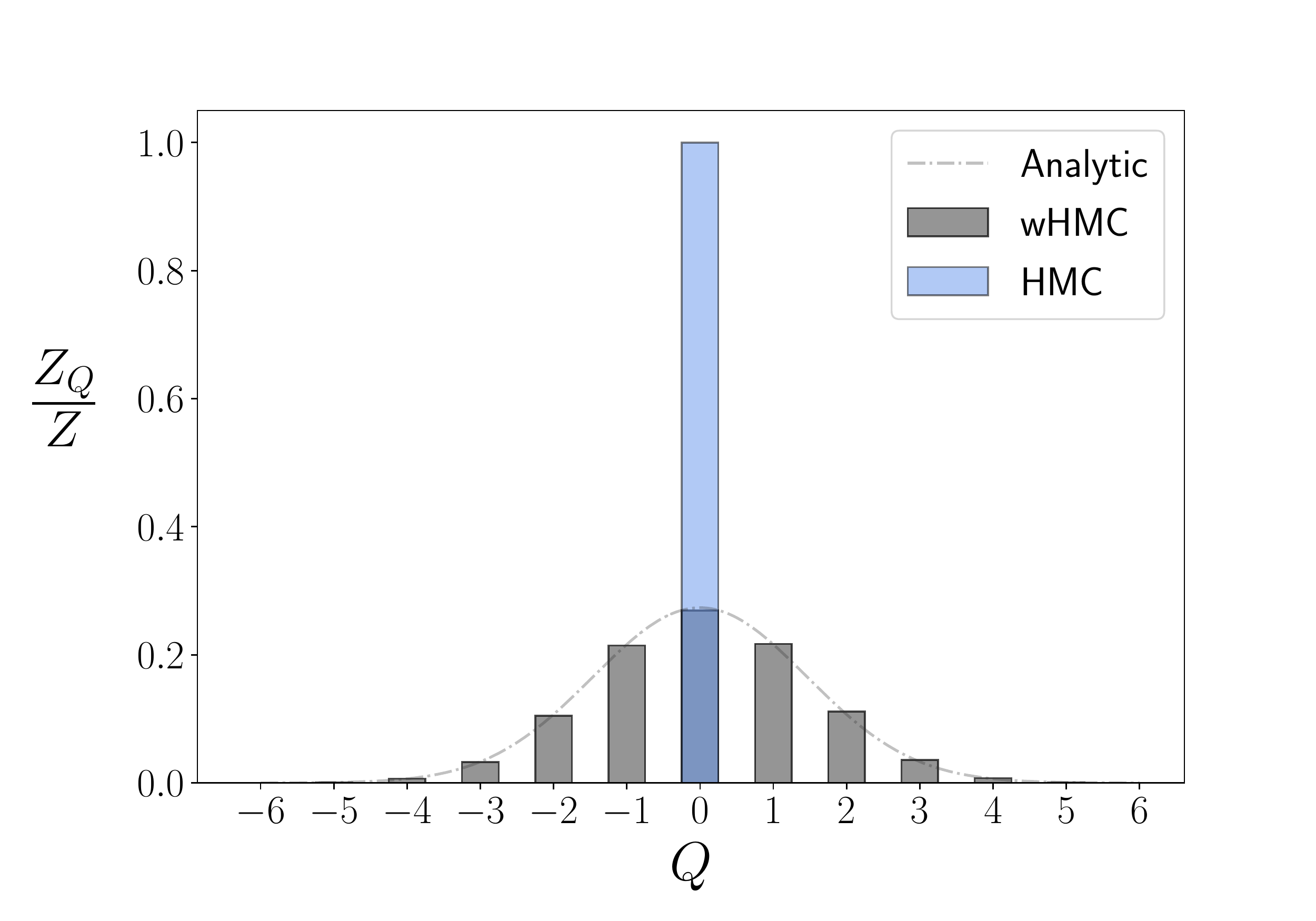}	
\caption{(Left) Monte Carlo history of the topological charge for the two
algorithms with $10^{6}$ configurations at $\beta=8.45$. (Right) Number of
configurations per topological sector for HMC and wHMC at $\beta=11.25$ for
$10^{6}$ configurations, along with the analytical result.}
	\label{fig:Qhistory}
\end{figure*}

Now that we can say that we have an algorithm with acceptance, the question to
ask is whether it performs better than HMC.  As shown in Fig. \ref{fig:Qhistory}
(left), in the pure gauge theory we ran two simulations, one with HMC and the
other with wHMC, and looked at the history of the topological charge $Q$: we can
see that we are at a $\beta$ where autocorrelations in HMC are noticeable, but
in wHMC are not.

Then, in Fig. \ref{fig:Qhistory} (right),  we increased $\beta$ and counted the
number of configurations in each topological sector, seeing that although HMC is
completely frozen in $Q=0$, winding HMC is able to sample from all the relevant
topological sectors reproducing the analytical result (dashed line). 

\begin{figure*}[t]
	\centering
\includegraphics[width=7.5cm,keepaspectratio]{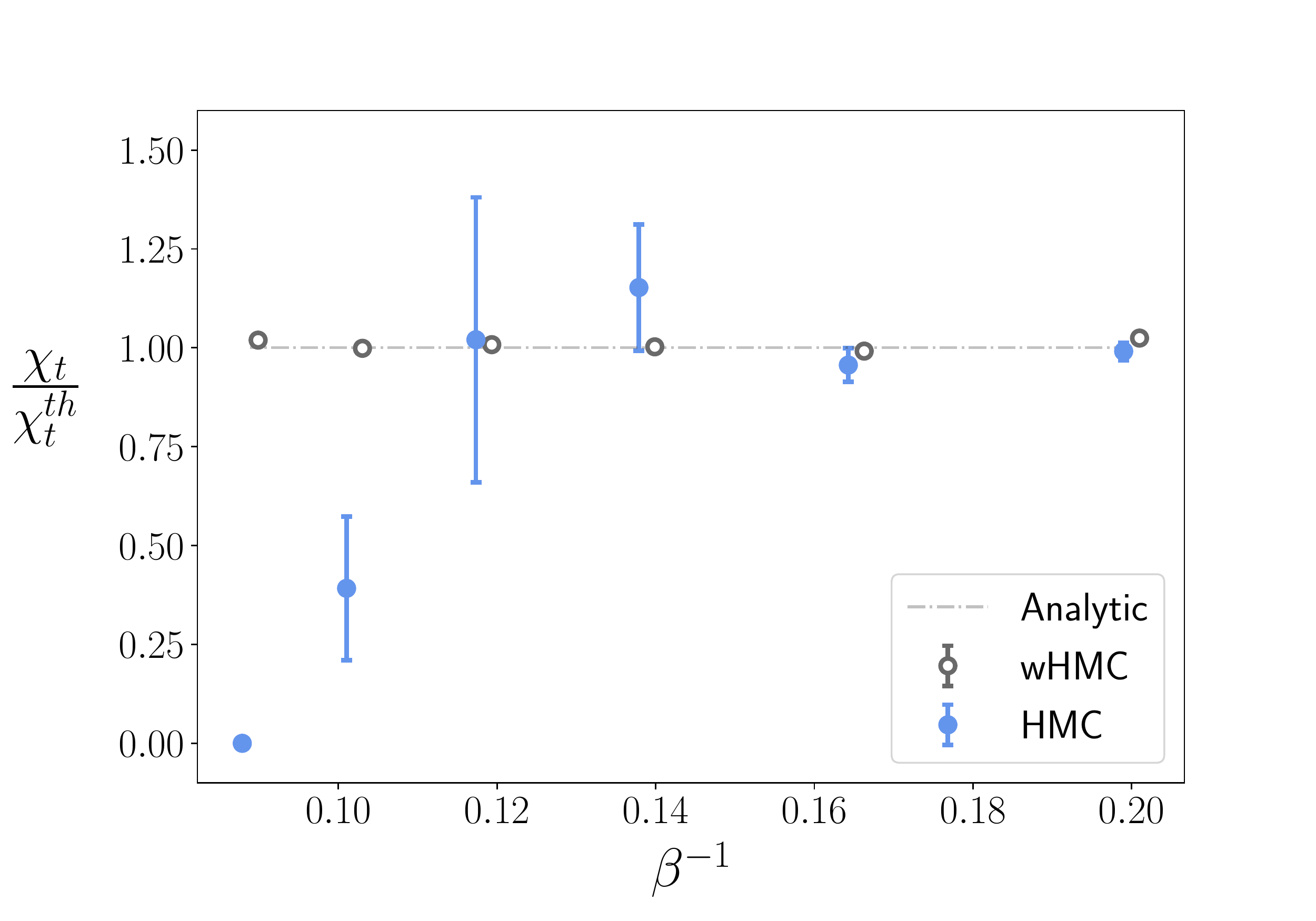}	
\includegraphics[width=7.5cm,keepaspectratio]{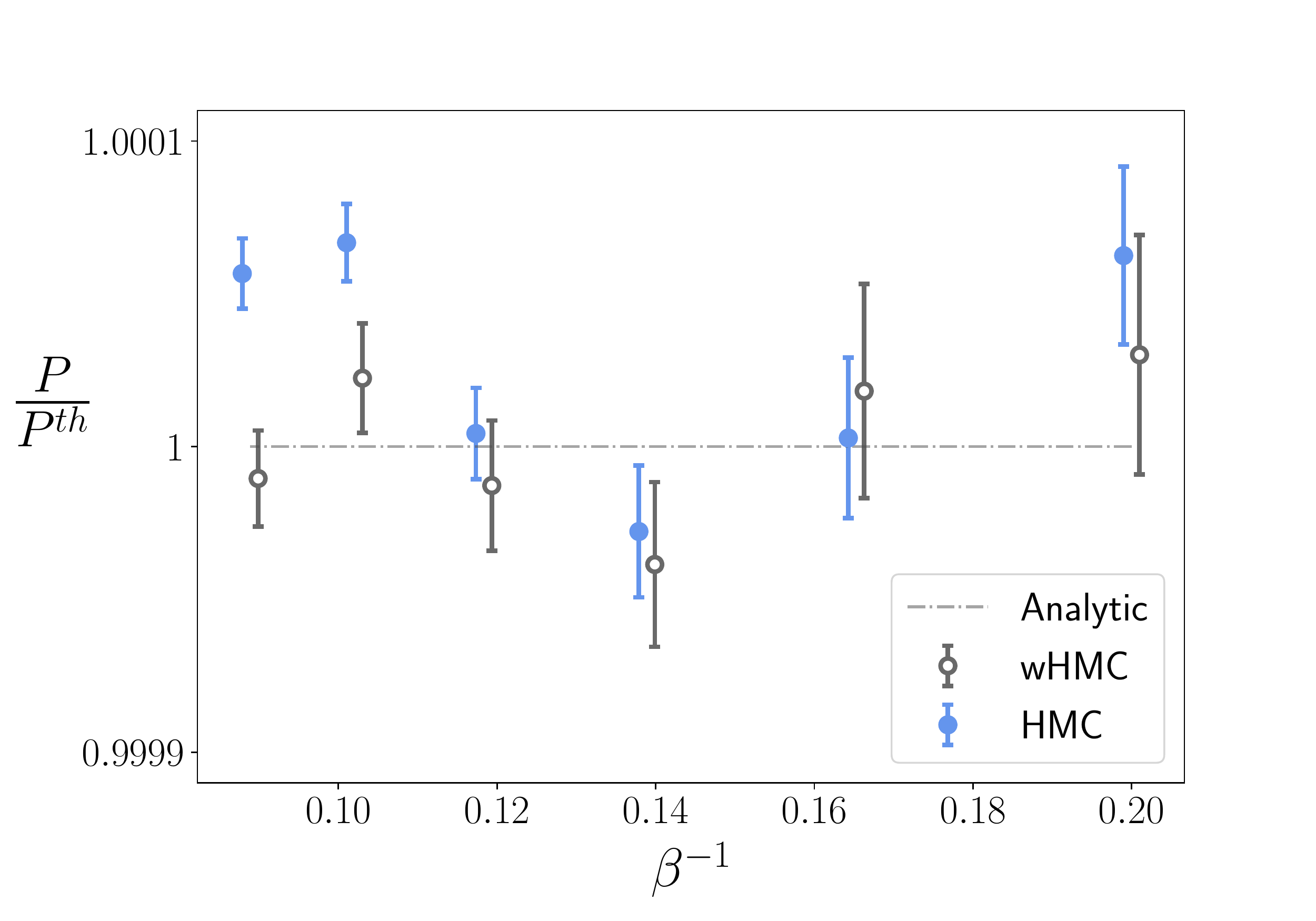}	
\caption{Average topological susceptibility (left) and plaquette (right)
normalized to the analytical result as a function of $\beta^{-1}$ for both
algorithms.}
	\label{fig:chi-P-vs-beta}
\end{figure*}

This implies that HMC would yield biased values for the observables we
obtain, while wHMC would give the correct ones, and indeed in the pure gauge
theory we can check directly with analytical results for all $\beta$ and volume
sizes.  For example, in Fig. \ref{fig:chi-P-vs-beta} (left) we plot the
topological susceptibility over the analytical one as a function of
$\beta^{-1}$, so we approach the continuum limit from right to left;  the
correct result would be 1, and we see that wHMC in gray gets the correct value
for all $\beta$, but HMC is clearly biased when we approach the continuum. 

But this behavior was expected because topology gets frozen in HMC, and the
topological susceptibility is a topological observable. The important point is
that even with the plaquette, shown in Fig. \ref{fig:chi-P-vs-beta} (right), which is a
non-topological observable, we observe the same behavior: HMC is also biased for
non-topological observables when topology is frozen.

\subsection{Fixed topology}

\begin{figure*}[t]
	\centering
\includegraphics[width=7.5cm,keepaspectratio]{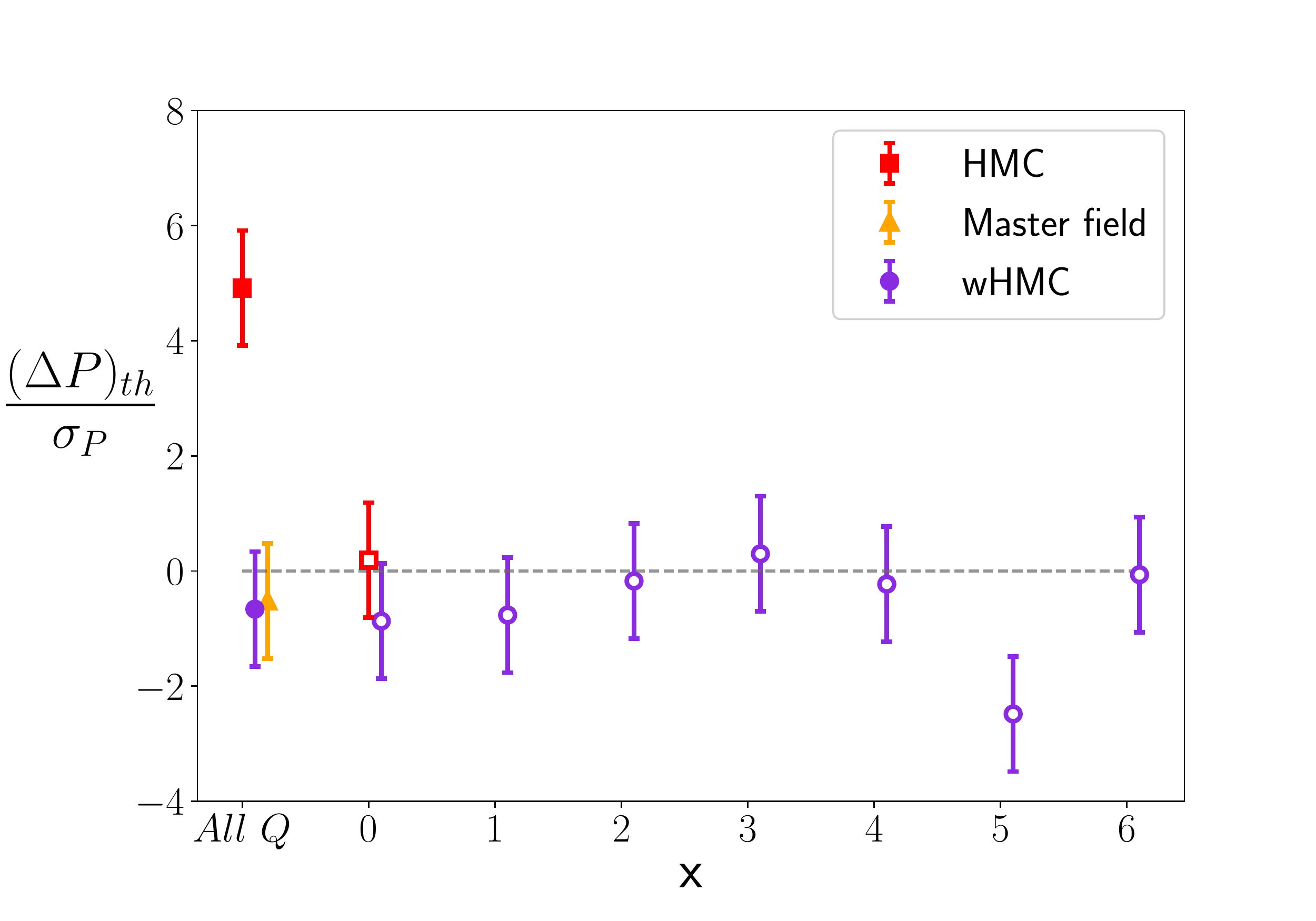}	
\includegraphics[width=7.5cm,keepaspectratio]{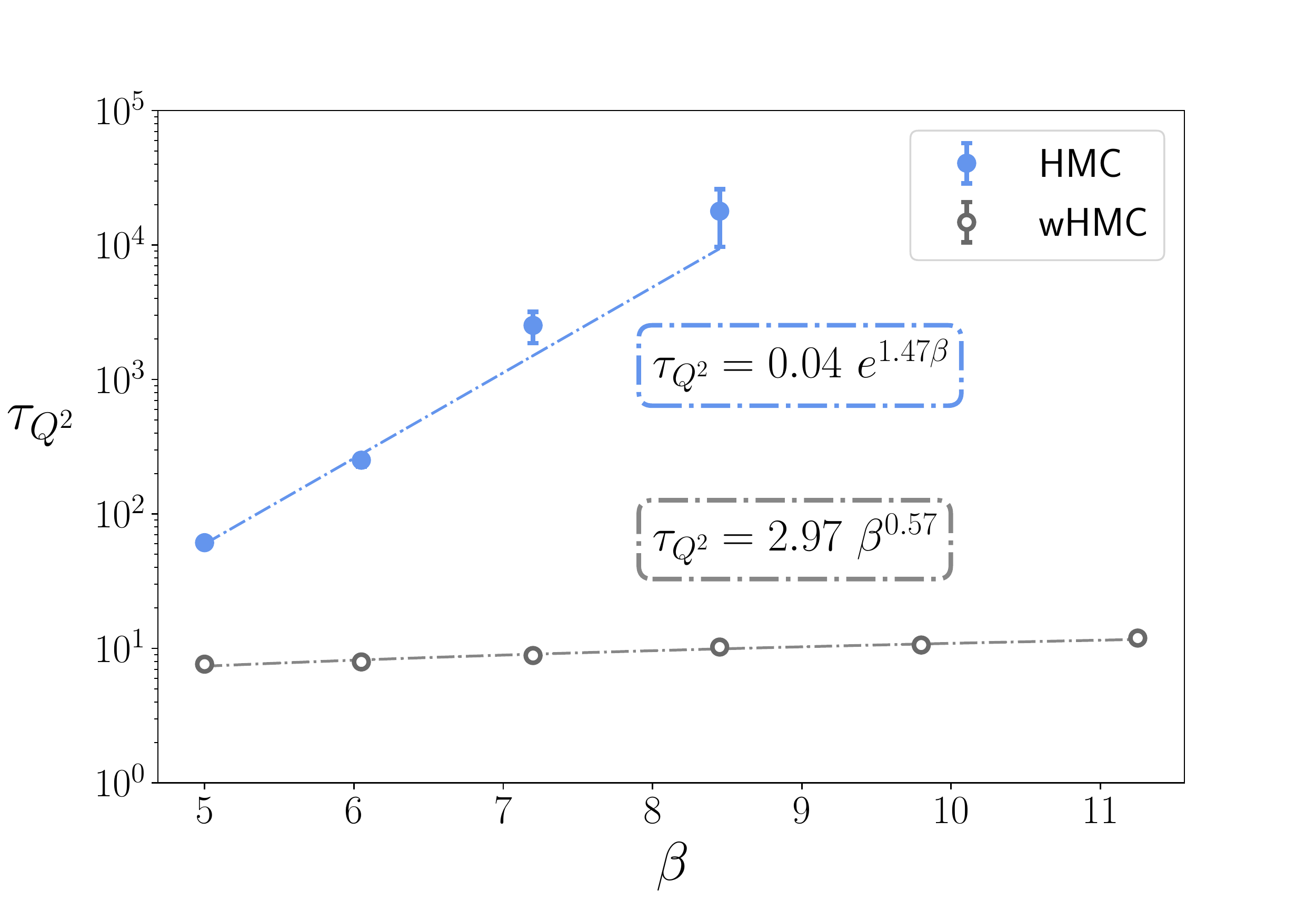}	
\caption{(Left) $(\Delta P)_{\text{th}} \equiv P_{\text{x}} -
	P_{\text{x}}^{\text{th}}$
normalized with the standard deviation $\sigma$ for $\text{x} = $ \emph{All}
$Q$ (full symbols) or at fixed topology $\text{x} = \left| Q \right|$ (open
symbols), at $\beta=11.25$ for wHMC, HMC and master field (see
\cite{Albandea_2021} for details), compared to the analytical result (dashed
line). (Right) Autocorrelation time of $Q^2$ as a function of $\beta \propto
a^{-2}$, as obtained with the wHMC and HMC algorithms.}
	\label{fig:P-disc}
\end{figure*}

Although yielding incorrect results when topology is frozen, there is a common
belief that HMC samples correctly the topological sectors in which it is frozen
(for instance, the sector $Q=0$ in Fig. \ref{fig:Qhistory}), which would imply
that there are no infinite action barriers separating different regions of the
same topological sector. One way to check this is by projecting our results to
the different topological sectors \cite{Albandea_2021}, by defining the
projected observable $O$ to the topological sector $n$ 
\begin{align}
	O_{n} = \frac{\left< O\, \delta_{n}(Q) \right>}{\left< \delta_{n}(Q)
	\right>}, \qquad \text{where }\,\,\,\,
	\delta_{n}(Q) = \begin{cases}
		1 \qquad \left| Q \right|=n \\
		0 \qquad \text{otherwise}
	\end{cases}.
\end{align}
In Fig. \ref{fig:P-disc} (left) we plot how many $\sigma$ discrepancy there is
from the analytical result (dashed line) for the plaquette $P$ of HMC and wHMC.
The first values in the plot are the average of the plaquette over all
topological sectors, which corresponds to the left-most values of Fig.
\ref{fig:chi-P-vs-beta} (right): wHMC in violet is consistent with the
analytical result, but HMC in red has a discrepancy of more than $4 \sigma$.

But the remarkable thing is that when we look at the results at the fixed
topological sector $\left| Q \right|=0$ we see that both wHMC and HMC are
consistent with the analytical result, even when topology in HMC is frozen,
indicating that HMC would indeed sample correctly within each topological
sector, no matter if topology is frozen or not.


\subsection{Scaling with the lattice spacing}

Finally, one is interested in how the autocorrelations of the algorithm
scale with the lattice spacing $a$. In Fig. \ref{fig:P-disc} (right) we plot the
autocorrelation of $Q^2$ as a function of $\beta \sim a^{-2}$ for HMC and wHMC,
and find that for HMC it increases exponentially, while wHMC increases just
polynomially with $\sim \sqrt{\beta}$, being an enormous improvement.

\section{$N_{f} = 2$ results}

\begin{figure*}[t]
	\centering
\includegraphics[width=7.5cm,keepaspectratio]{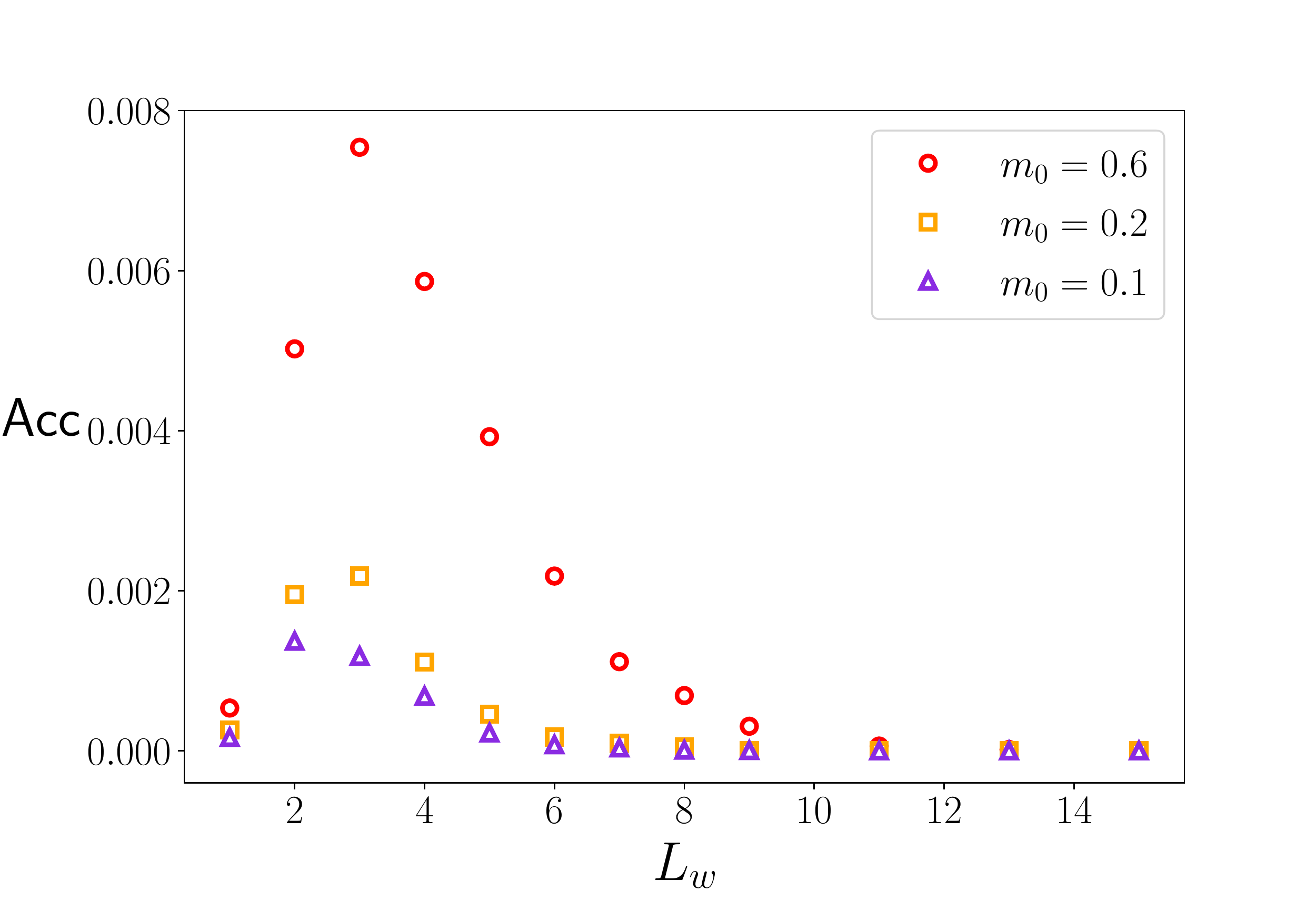}	
\includegraphics[width=7.5cm,keepaspectratio]{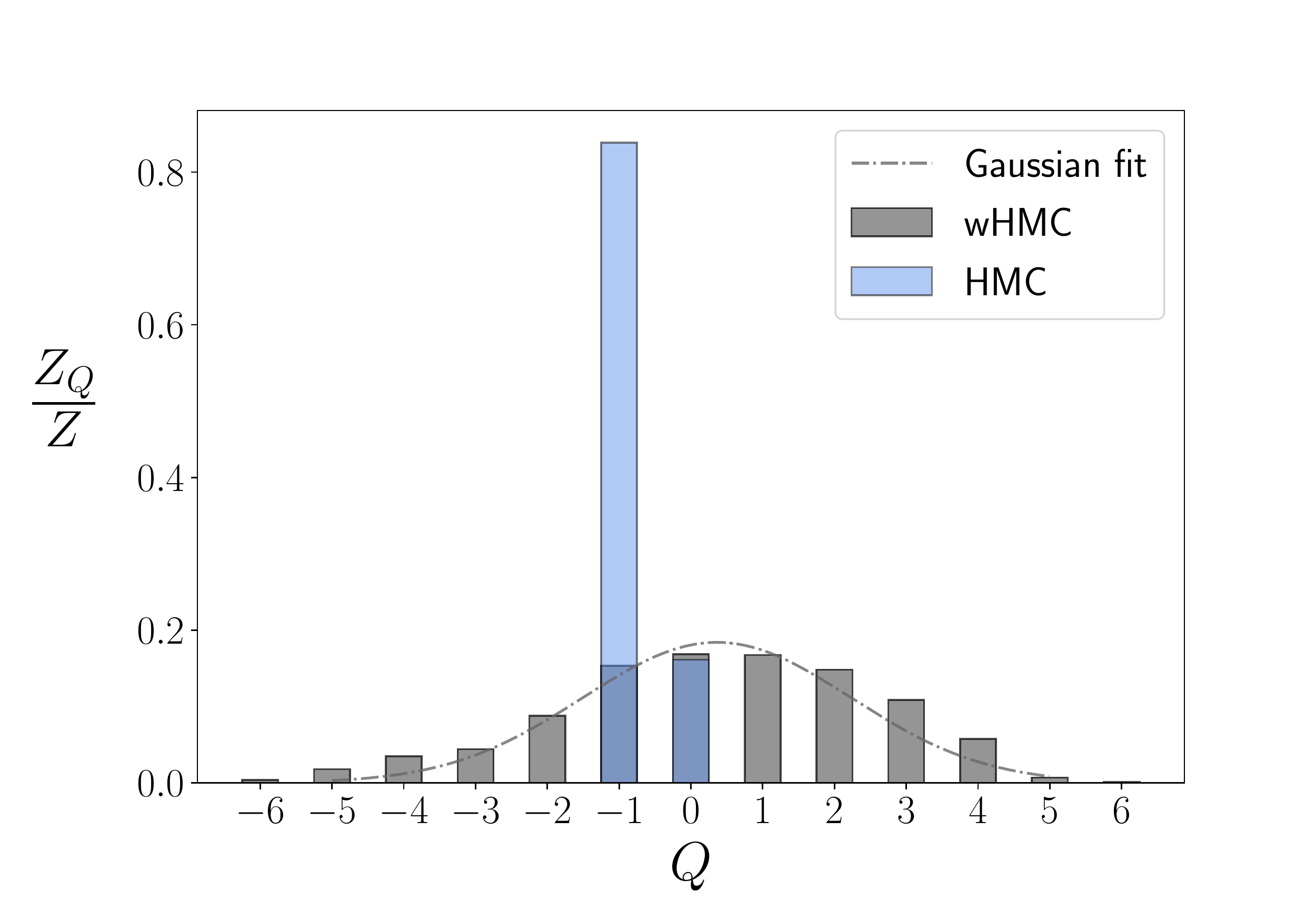}	
\caption{(Left) Acceptance of a winding step in wHMC as a function of $L_{w}$
for various bare quark masses at $\beta=5.0$. (Right) Number of configurations
per topological sector at $\beta=9.0$ for HMC and wHMC with $N_{f}=2$. A
Gaussian fit to the wHMC distribution is also shown.}
	\label{fig:facc}
\end{figure*}

Adding two dynamical fermions we additionally have to sample the determinant of
the Dirac operator, which is a highly non-local object. This is a problem
because our winding is a transformation of the gauge links, and this will change the
fermionic action; therefore, we expect that the acceptance of the algorithm will
decrease. In Fig. \ref{fig:facc} (left) we show the acceptance of the winding
step versus $L_{w}$ for three different bare masses of the quark, and indeed now
we are talking about acceptances that are much lower than the ones in the pure
gauge theory (cf. Fig. \ref{fig:winding_trafo}). Also, we see that now there is
an optimal size $L_{w}$ for the winding transformation.

A way to alleviate this problem is to perform more winding transformations per
step, balancing up the computational time devoted to HMC evolutions and to
winding transformations. The structure of the algorithm would now be:
\begin{enumerate}
	\item
		Perform a molecular dynamics evolution using HMC.
	\item
		Accept or reject the new configuration using Eq.
		\eqref{eq:MHacc}.
	\item
		Perform a winding or antiwinding transformation. 
		\label{it:item1}
	\item
		Accept or reject the new configuration using Eq.
		\eqref{eq:MHacc}.
		\label{it:item2}
	\item
		Repeat steps \ref{it:item1}--\ref{it:item2} $N_{w}$ times.
	\item
		Repeat.
\end{enumerate}
This would now be a wHMC step, where $N_{w}$ is the number of winding
transformations performed per HMC evolution and is a parameter to be set before
the simulation to optimize the probability of making a transition of topological
sector.

In Fig. \ref{fig:facc} (right) we check that, at equivalent computational costs,
wHMC is able to sample from all the relevant topological sectors while HMC is
not, so again one expects that HMC will lead to biased results for the
observables.

\begin{figure*}[t]
	\centering
\includegraphics[width=10cm,keepaspectratio]{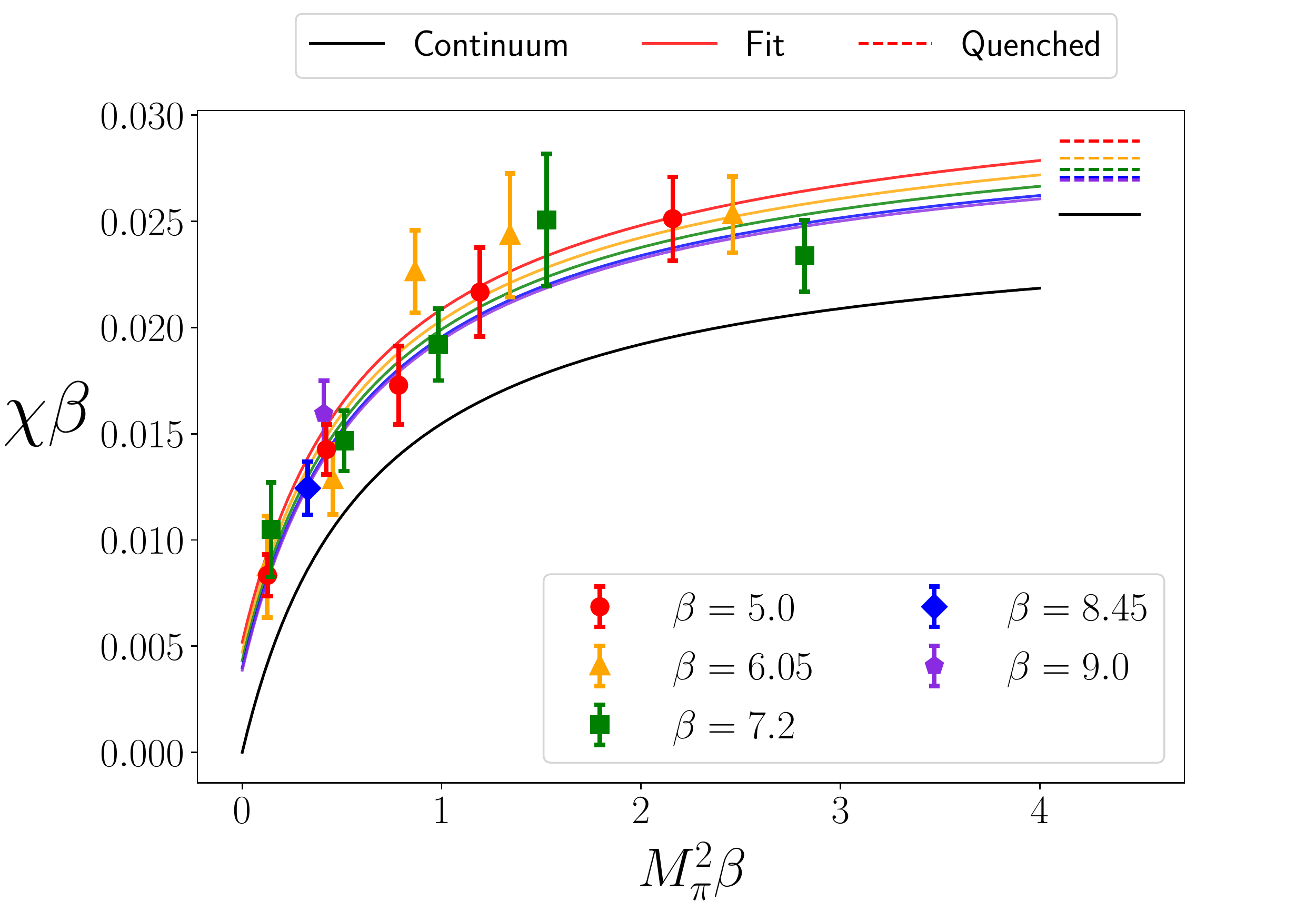}	
\caption{Topological susceptibility in the $N_{f}=2$ theory as a function of the
pion mass $M_{\pi}$. The coloured solid lines are fits to the expression in
Eq. \eqref{eq:chi-M-cutoff} for five $\beta $'s, while the black line is the
continuum result. The horizontal dashed lines are the quenched expectation at
the various $\beta$ and the continuum.}
	\label{fig:chi-vs-M}
\end{figure*}

Also, close to the chiral limit one can derive the relation \cite{Albandea_2021}
\begin{align}
	\chi_{t}^{N_{f}} = \frac{1}{4\pi\beta } \frac{M_{\pi}^{2} \beta }{N_{f}
	+ \pi M_{\pi}^{2}\beta }
	\label{eq:chi-vs-M}
\end{align}
between the topological susceptibility and the mass of the pion, which nicely
interpolates between the quenched topological susceptibility
$X_{t}|_{\text{quenched}} = 1 / 4\pi^2\beta $ for $M_{\pi} \rightarrow \infty$,
and the chiral limit $M_{\pi} \rightarrow 0$, where the topological
susceptibility vanishes.

In Fig. \ref{fig:chi-vs-M} we show the topological susceptibility as a function
of the pion mass, together with the fit to the continuum expectation, Eq.
\eqref{eq:chi-vs-M}, plus generic cutoff effects,
\begin{align}
	\chi_{t}^{N_{f}=2} = \text{Eq.}(\ref{eq:chi-vs-M}) + \left( c +
	dM_{\pi}^{2} \right) \beta^{-1 / 2},
	\label{eq:chi-M-cutoff}
\end{align}
with $c$ and $d$ fitting parameters. The agreement of wHMC with the expectation
is good, even at values of $\beta$ where the topology in HMC is completely
frozen and does not allow to measure the topological susceptibility.

\subsection{Fixed topology}

\begin{figure*}[t]
	\centering
\includegraphics[width=7.5cm,keepaspectratio]{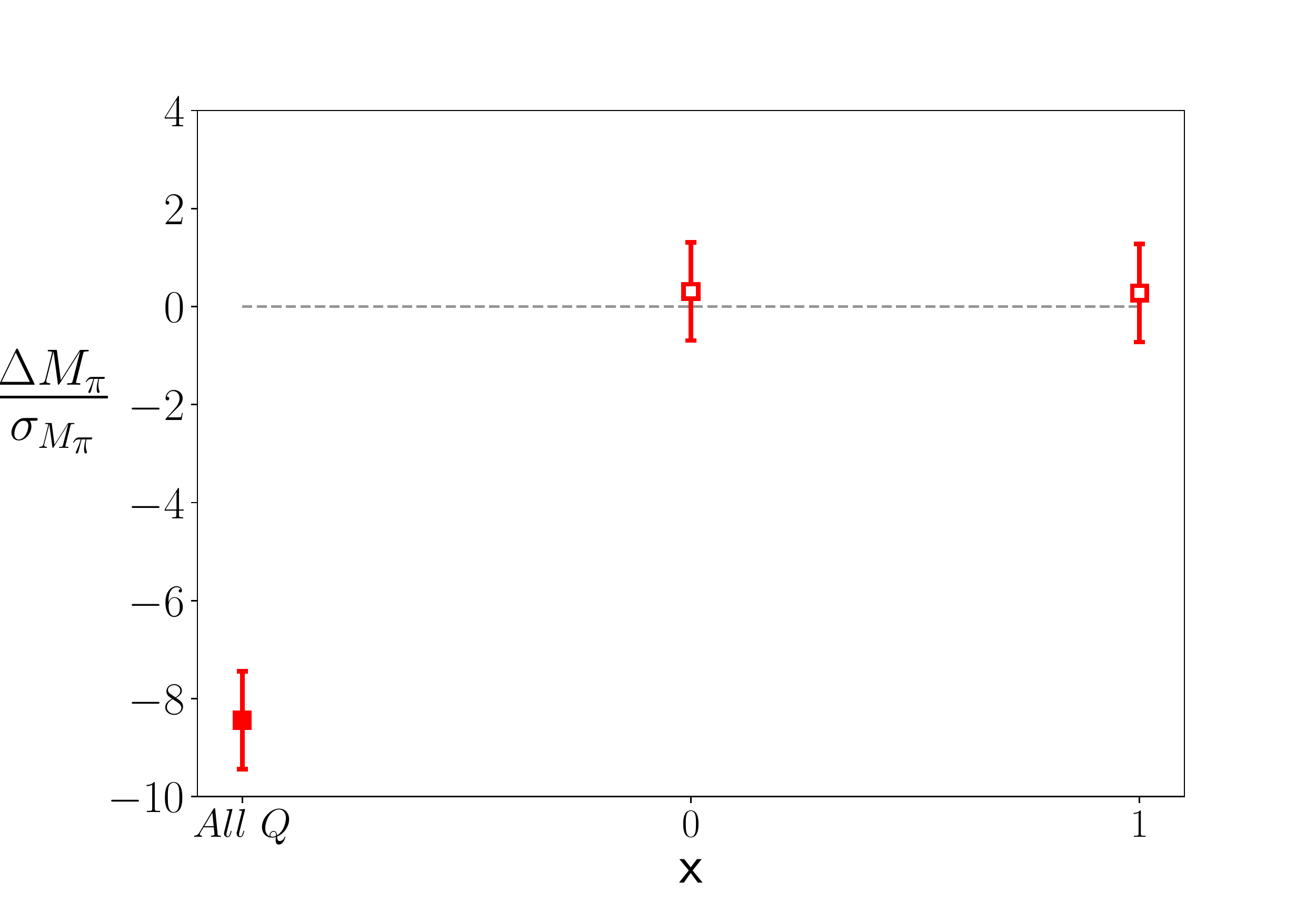}	
\includegraphics[width=7.5cm,keepaspectratio]{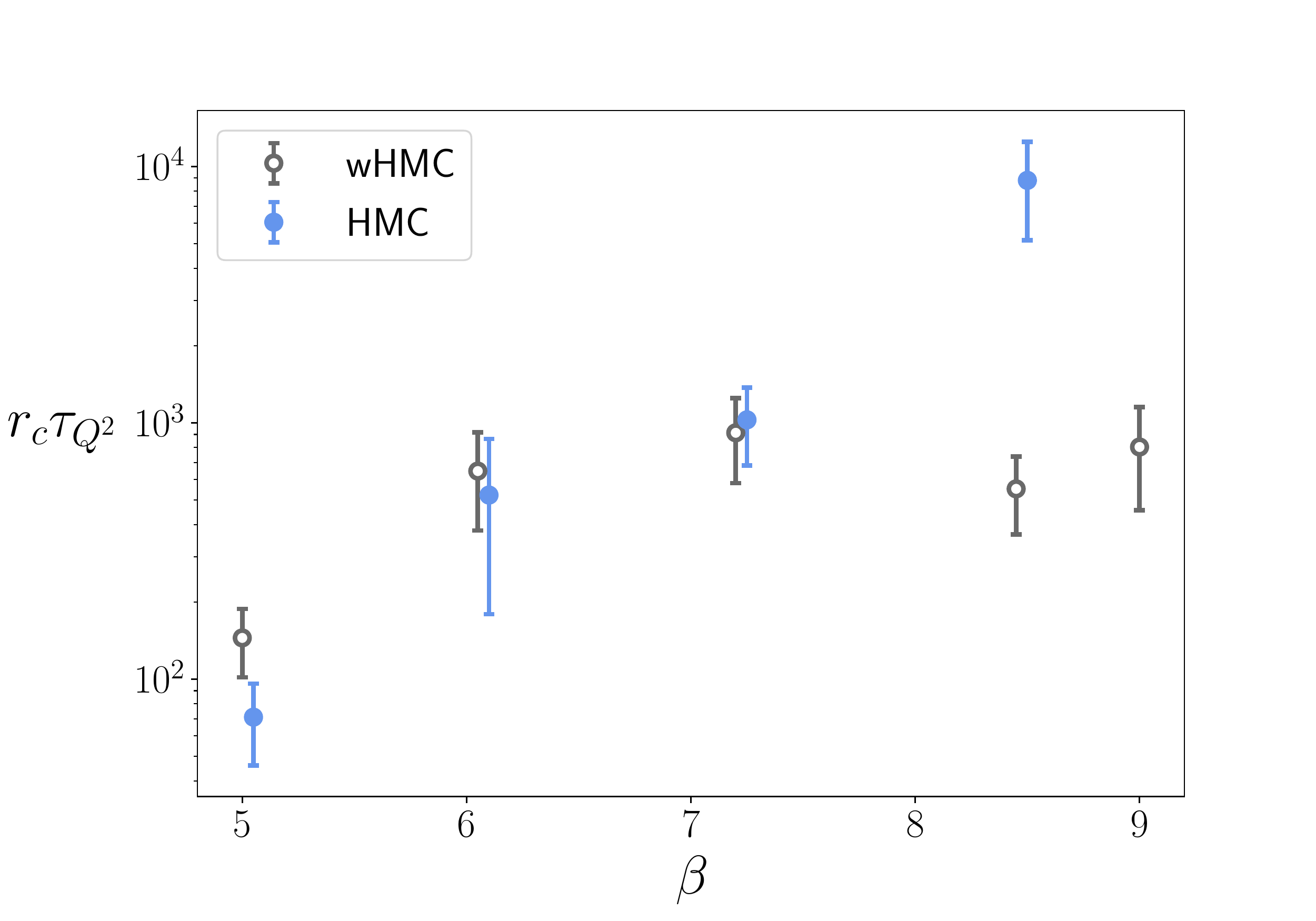}	
\caption{(Left) $\Delta M_{\pi} \equiv \left< M_{\pi} \right>_{\text{x}}^{\text{HMC}} -
\left< M_{\pi} \right>_{\text{x}}^{\text{wHMC}}$, averaged over all sectors,
$\text{x} = $ \emph{All} $Q$ (full symbol), or sectors of fixed $\text{x} =
\left| Q \right|$ (open symbols), at  $\beta=9.0$. (Right) Scaling of the
autocorrelation time for $Q^2$ with $\beta $ for HMC and wHMC. The factor
$r_{c}$ accounts for the differences in computational cost of the two
algorithms. The pion mass is kept approximately constant, $M_{\pi} \sqrt{\beta }
\sim 0.65$.}
	\label{fig:M-disc}
\end{figure*}

With fermions there are no analytical results, but we can look at the
discrepancy between both algorithms for the mass of the pion $M_{\pi}$. In Fig.
\ref{fig:M-disc} (left) we plot the discrepancy of $M_{\pi}$ between the two
algorithms versus the topological sector, and we can see that we have the same
behavior as in the pure gauge theory: for the topological average, \emph{All
$Q$}, HMC has an $8\sigma$ discrepancy with respect to wHMC, despite the mass is
a non-topological quantity; nonetheless, both algorithms agree in the fixed
topological sectors $\left| Q \right|=0$ and $\left| Q \right| = 1$. This is
another indication that HMC samples correctly at a fixed topological sector,
despite topology being frozen.

\subsection{Scaling with the lattice spacing}

Finally, in Fig. \ref{fig:M-disc} (right) we look at the scaling of the
autocorrelation of $Q^2$ as a function of $\beta \sim a^{-2}$ for both HMC and
wHMC at equivalent computational cost, and we see that, while the scaling of HMC
is still exponential, we no longer have the nice polynomial scaling for wHMC
that we had in the pure gauge theory. However, we still have a significant
improvement in the scaling towards the continuum limit.

\section{Outlook}

We have presented a new algorithm based on Metropolis--Hastings steps that are
tailored to induce jumps in the topological charge. This algorithm satisfies
detailed balance, and ergodicity is ensured when alternated with standard HMC
steps. As we have shown, it successfully improves the problem of topology
freezing and exponentially-growing autocorrelation times in the 2D model
considered---both with and without fermion content. Also, we
have been able to confirm that averages in fixed topology sectors are not
affected by topology freezing, and agree in wHMC and HMC. This is seen both in
the pure gauge theory, where the analytical results are known at finite $\beta$,
as well as in the theory with fermions.

The interesting question is whether wHMC can be equally successful in the case
of other gauge theories in higher dimensions. In fact, the winding step is
trivial to extend to, for instance, a SU(2) theory in 4D. We have indeed carried
out the naive implementation of wHMC in that context, and found very poor
acceptances---the ``curse'' of dimensionality. We hope that less trivial
implementations in 4D could resolve this matter.

\begin{acknowledgments}
We acknowledge support from the Generalitat Valenciana grant PROMETEO/2019/083,
the European project H2020-MSCA-ITN-2019//860881-HIDDeN, and the national
project FPA2017-85985-P. AR and FRL acknowledge financial support from
Generalitat Valenciana through the plan GenT program (CIDEGENT/2019/040). DA
acknowledges support from the Generalitat Valenciana grant ACIF/2020/011. The
work of FRL is supported in part by the U.S. Department of Energy, Office of
Science, Office of Nuclear Physics, under grant Contract Numbers DE-SC0011090
and DE-SC0021006. We acknowledge the computational resources provided by Finis
Terrae II (CESGA), Lluis Vives (UV), Tirant III (UV). The authors also
gratefully acknowledge the computer resources at Artemisa, funded by the
European Union ERDF and Comunitat Valenciana, as well as the technical support
provided by the Instituto de Física Corpuscular, IFIC (CSIC-UV).  
\end{acknowledgments}

\end{document}